\documentclass{iau}

\usepackage{amsmath}
\usepackage{graphicx}
\usepackage{multirow}

\begin{document}

\lefttitle{Rivilla}
\righttitle{First interstellar detections of molecules towards G+0.693}

\jnlPage{1}{7}
\jnlDoiYr{2026}
\doival{}

\aopheadtitle{Proceedings IAU Symposium}
\editors{M. Zaja\v{c}ek,  T. Je\v{r}\'{a}bkov\'{a}, V. Karas, R. Schödel \&  P. Sukov\'{a}, eds.}

\title{The Central Molecular Zone of our Galaxy, a chemical complexity factory: first interstellar detections of molecules towards G+0.693}

\author{Víctor M. Rivilla}
\affiliation{Centro de Astrobiolog{\'i}a (CAB), CSIC-INTA, Carretera de Ajalvir km 4, Torrej{\'o}n de Ardoz, 28850, Madrid, Spain}

\begin{abstract}

The Central Molecular Zone (CMZ) of our Galaxy is a prime laboratory for interstellar chemistry because of its extreme conditions, such as shocks and high cosmic-ray ionisation rates. It is an efficient factory of molecules, with many species first discovered in CMZ sources. This document reviews key results from the study of the CMZ G+0.693–0.027 molecular cloud. Using ultra-sensitive broadband spectral surveys together with laboratory spectroscopy, more than 140 molecules have been identified, including 28 detected for the first time in the interstellar medium (ISM). These discoveries include prebiotic precursors of biomolecules, increasingly complex members of known molecular families, molecules containing periodic third-row elements (S, P, Na, Mg), high-energy stereo- and structural isomers, and new cations. Overall, these findings extend the known limits of interstellar chemical complexity and highlight the CMZ’s potential to provide the molecular ingredients for prebiotic chemistry.

\end{abstract}


\maketitle

%\section{Introduction}

\begin{figure*}[h]
\centering
\includegraphics[width=0.98\textwidth]{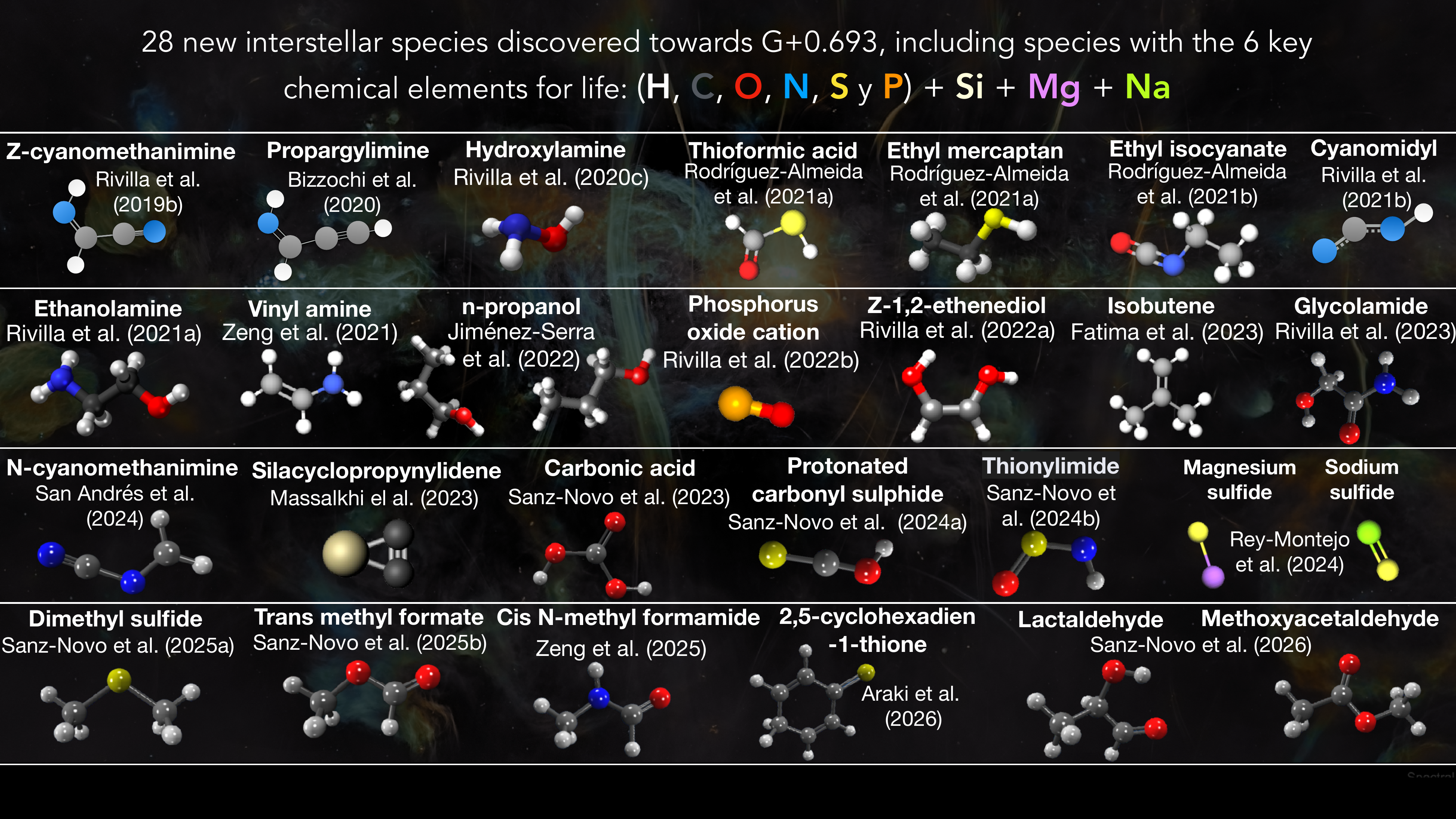}
\vspace{-3mm}
\caption{Molecules identified for the first time in the interstellar medium (ISM) towards the G+0.693-0.027 molecular cloud. The names of the species, the works reporting the detections, and their 3-dimensional representations are shown. White, gray, red, blue, yellow, orange, brown, purple and green balls represent atoms of hydrogen (H), carbon (C), oxygen (O), nitrogen (N), sulfur (S), phosphorus (P), silicon (Si), magnesium (Mg), and sodium (Na), respectively. Some of the 3-dimensional representations are courtesy of M. Sanz-Novo.} 
\label{fig:new-detections} 
\end{figure*}

\clearpage

\begin{figure*}
\centering
\includegraphics[width=\textwidth]{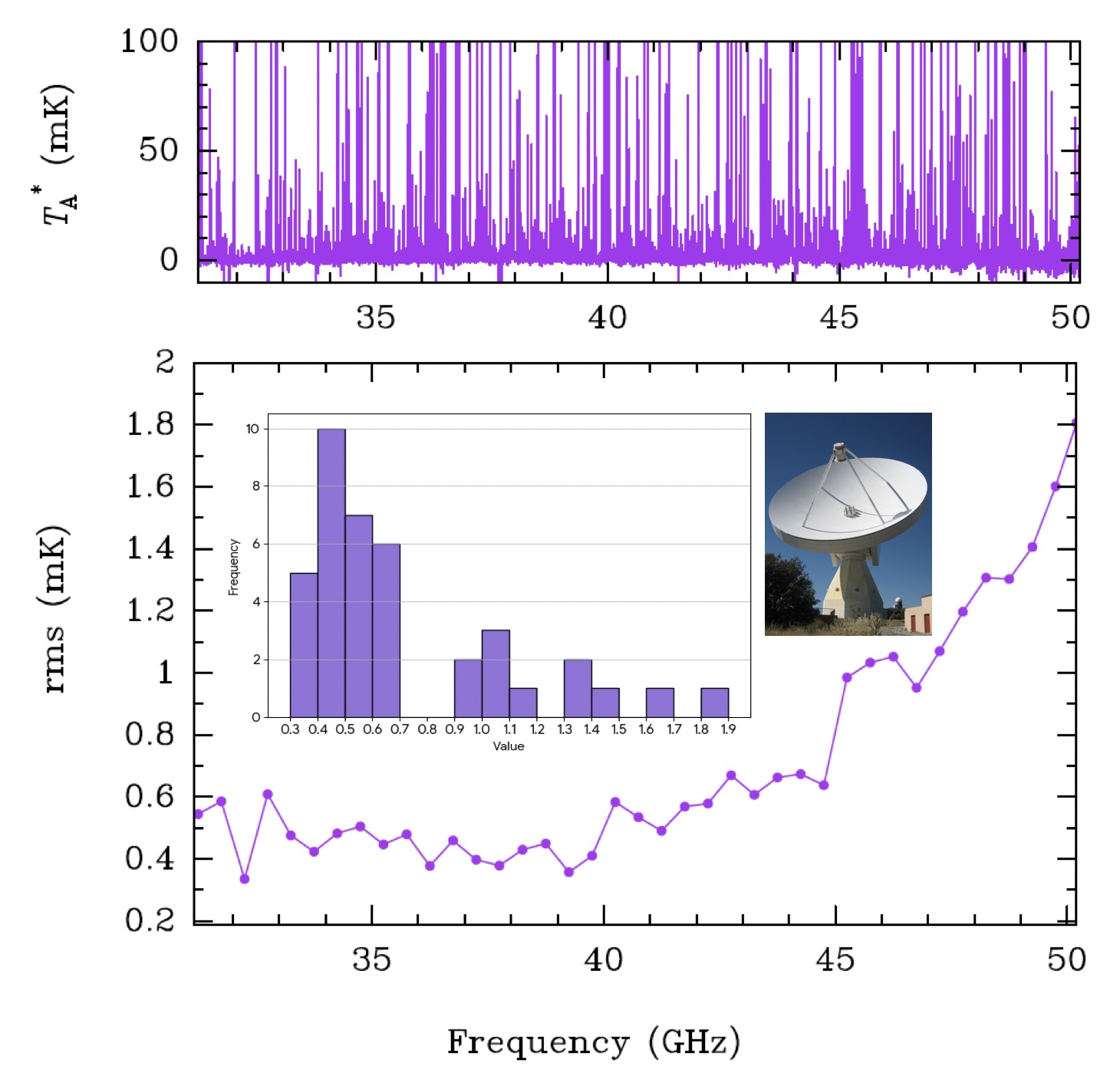}
\vspace{-8mm}
\caption{{\it Upper panel}: Spectral survey towards the G+0.693 molecular cloud obtained in the Q-band with the Yebes 40m telescope (Guadalajara, Spain). {\it Lower panel:} Values of the $rms$ noise of the spectra as a function of frequency. The left inset shows the histogram distribution of the $rms$, and the right inset shows a picture of the Yebes 40m telescope.} 
\label{fig:yebes-survey} 
\end{figure*}

The Central Molecular Zone (CMZ) of our Galaxy harbors one of the most chemically rich Galactic sources: the G+0.693-0.027 molecular cloud (hereafter G+0.693). It is located in the Sgr B2 complex, $\sim$55" north-east from the Sgr B2(N) massive protocluster (see e.g., Fig. 10 of \citealt{colzi2024}). Its physical and chemical properties make it an excellent astrochemical laboratory, and a well-suited target for the discovery of new interstellar species. Although there is no sign of ongoing star formation, it is thought that the region might produce the next generation of protoclusters in the Sgr B2 region (\citealt{colzi2024}). The intermediate gas densities, which span the range 10$^4$ -10$^5$ cm$^{-3}$ (\citealt{Zeng2020,Colzi2022,colzi2024}), produce sub-thermal excitation of the molecules. This translates into low excitation temperatures ($T_{\rm ex}$) of 5$-$20 K (e.g., \citealt{Requena-Torres2008,Zeng2018,Rivilla2022_nitriles}), well below the typical high kinetic temperatures of CMZ clouds (e.g., \citealt{Krieger2017}). As a consequence, only the lower energy levels of the molecules are excited, producing a much less crowded spectrum than in sources with higher $T_{\rm ex}$, such as hot cores and corinos. Therefore, line contamination is much less severe, which facilitates the detection of weak lines of new interstellar species. Moreover, G+0.693 harbors large-scale shocks, which are thought to be produced by a cloud-cloud collision event (\citealt{Zeng2020,colzi2024}). These shocks are able to sputter the dust grains, and to inject into the gas phase the molecular species formed on their icy mantles. Furthermore, the high cosmic-ray ionisation rates present in the CMZ, and in particular in G+0.693 (\citealt{Rivilla2022_PO+,Sanz-Novo2024a}), which are orders of magnitude higher that those found in clouds in the Galactic disk, are expected to boost the formation of radicals in the dust surfaces. This can trigger radical-radical reactions that form species of increasing molecular diversity and size (\citealt{Rivilla2023}). From all this, G+0.693 is considered a unique chemical complexity factory.

%
%It can be formed through radical-radical reactions on the surface of dust grains. The abundances of these radicals can be significantly boosted in an environment affected by a strong ultraviolet field induced by cosmic rays, such as that expected in G+0.693-0.027.
%
% ~55″ north-east from Sgr B2(N)
%
% cloud-cloud collision
%\citealt{Zeng2020}
%
% densities of 10^4 -10^5
%\citealt{Zeng2020,Colzi2022,colzi2024}
%
% low excitation temperatures 
%\citealt{Requena-Torres2008,Zeng2018,Rivilla2022_nitriles}
%
% lower than the typical temperatures of CMZ clouds.
%\citealt{Krieger2017}
%
% low fust temperatures ~20 K
%\citealt{battersby2025}
%
% high cosmic-ray ionisation rate 
%\citealt{Rivilla2022_PO+,Sanz-Novo2024a}
%
% prestellar cluster stage
%\citealt{colzi2024}
%
% radical-radical reactions in the surgace of dust grains
%\citealt{Rivilla2023}
%

\begin{figure*}
\centering
\includegraphics[width=\textwidth]{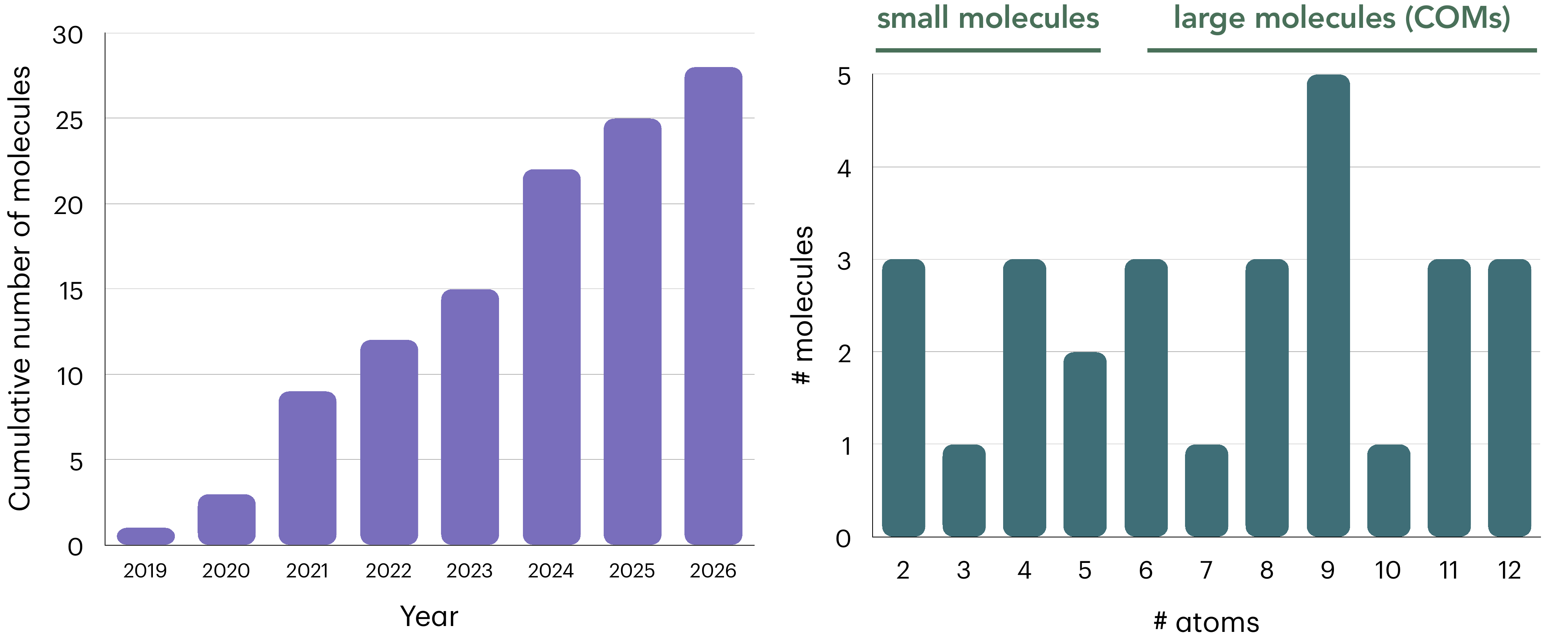}
\vspace{-7mm}
\caption{{\it Left panel}: Cumulative number of molecular species firstly discovered in the ISM towards the G+0.693 molecular cloud in the period 2019-2026 (until the celebration of IAUS 405 on May 2026). {\it Right panel}: Distribution of the molecules detected for the first time in the ISM towards the G+0.693 molecular cloud as a function of the number of constituent atoms: 9 small ($<$6 atoms) species, and 19 large species ($\geq$6 atoms), commonly known as complex organic molecules (COMs; \citealt{vD&herbst2009}).} 
% 9 small molecules, and 19 COMs 
\label{fig:histo} 
\end{figure*}

Indeed, 
%a highly sensitive spectral survey combining data from different telescopes (Yebes 40m - see Fig. \ref{fig:yebes-survey}, IRAM 30m, GBT and APEX) has allowed the first detections in the ISM of 
28 new molecules have been first detected in the ISM towards G+0.693 in the last 8 years (see Fig. \ref{fig:new-detections}). These discoveries have been possible thanks to the synergy between ultra-sensitive spectral surveys combining data from different telescopes (Yebes 40m - see Fig. \ref{fig:yebes-survey}, IRAM 30m, GBT and APEX), and new spectroscopic work in the laboratory (e.g., \citealt{Bizzocchi2020,Melosso2022,Fatima2023,Araki2026}).  
Table \ref{tab:detections} shows the list of molecules, indicating their names, chemical formulae, year of detection, number of constituent atoms, derived column densities, and literature works reporting them.
The left panel of Fig. \ref{fig:histo} shows the cumulative number of detections in the last years until the celebration of the IAUS 405 on May 2026. 
These discoveries rank G+0.693 as the third interstellar source where more  molecules have been discovered, already overcoming the Orion KL hot core, and only after the TMC-1 dark cloud and Sgr B2 - considering the total number of detections towards Sgr B2(N), and Sgr B2(M), Sgr B2(S), and Sgr B2(OH) - (\citealt{McGuire2022}).
The detected species contain the six key elements for life: Hydrogen [24 molecules], Carbon [24], Oxygen [16], Nitrogen [12], Sulfur [6] and Phosphorus [1]; and also other elements such as Silicon [1], Magnesium [1], and Sodium [1] (see Fig. \ref{fig:new-detections}). Among these species, 9 are small molecules (considered as those with $<$ 6 atoms), and 19 have $\geq$ 6 atoms (see Fig. \ref{fig:histo} right), which are commonly known as complex organic molecules (COMs; \citealt{vD&herbst2009}).
These newly identified interstellar species include prebiotic precursors of biomolecules, increasingly complex members of known molecular families, high-energy stereo- and structural isomers, and new cations (see details in the references listed in Table \ref{tab:detections}). Overall, these findings extend the known limits of interstellar chemical complexity, and highlight the CMZ’s potential to provide the molecular ingredients for prebiotic chemistry.
%
% relevance to prebiotic chemistry
%\citealt{Jimenez-Serra2020,Rivilla2022_nitriles}
%
%Statistics about C, N, O, S, P, and also Si, Mg, Na.
%
% H   C	 O	N	S	Si	P	Na	Mg
% 24 24	16	12	6	1	1	1	1
%
%The Central Molecular Zone (CMZ) of our Galaxy is an excellent laboratory to study interstellar chemistry due to its extreme physical conditions (e.g.,  shocks or high cosmic-ray ionisation rates…).  Indeed, the CMZ is a highly productive factory of interstellar molecules, and many of the species detected in the interstellar medium were first discovered in CMZ sources. In this talk I will review the key findings of the exploration of the chemical complexity of the CMZ G+0.693-0.027 molecular cloud. By combining ultra-sensitive broadband spectral surveys with spectroscopic laboratory studies, our group has identified over 140 molecules to date, including 28 species detected for the first time in the interstellar medium. These include: i) prebiotic precursors of RNA nucleotides, proteins, lipids and sugars; ii) increasingly complex molecules belonging to known interstellar families (amides, imines, carboxylic acids, alcohols, amines, thiols);  iii) new molecules containing chemical elements of the third row of the periodic table (S, P, Na, Mg); iv) new (high-energy) members of (stereo)isomeric families; and v) new interstellar cations. These results continuously push the known limits of interstellar complexity, and highlight its potential to produce the chemical feedstock necessary for the initiation of prebiotic chemistry.

\begin{table}
\centering
\tabcolsep 4pt 
 \caption{First detections of interstellar molecules discovered towards the G+0.693-0.027 molecular cloud$^{\rm [a]}$.}\label{sample-table}
 {\tablefont\begin{tabular}{@{\extracolsep{\fill}}ccccc c}
    \midrule
    Name & Formula & Year & Number  & Column density  & Reference \\
     &  & of detection  &  of atoms &   $N$$^{\rm [b]}$ (cm$^{-2}$) &  \\
    \midrule
C-cyanomethanimine$^{\rm [c]}$  & Z-HNCHCN & 2019 & 6 &  7.5(0.3) $\times$10$^{13}$ & \citet{Rivilla2019} \\
Propargylimine    & Z-NHCHCCH & 2020 & 7 &   2.40(0.02) $\times$10$^{13}$ & \citet{Bizzocchi2020} \\
Hydroxylamine    & NH$_2$OH & 2020 & 5 &   2.8(0.6) $\times$10$^{13}$ & \citet{Rivilla2020b} \\
Thioformic acid  & t-HCOSH  & 2021 & 5 &   1.6(0.1) $\times$10$^{13}$ & \citet{Rodriguez-Almeida2021a} \\
Ethyl mercaptan    & C$_2$H$_5$SH  & 2021   & 9 & 4(2) $\times$10$^{13}$ & \citet{Rodriguez-Almeida2021a} \\
Ethyl isocyanate    & C$_2$H$_5$NCO & 2021   & 10 & 8.1(1.3) $\times$10$^{12}$ & \citet{Rodriguez-Almeida2021b} \\
Cyanomidyl radical    & HNCN & 2021   & 4 & 1.1(0.1) $\times$10$^{13}$ & \citet{Rivilla2021b} \\
Ethanolamine & NH$_2$CH$_2$CH$_2$OH & 2021 & 11 &  1.51(0.07) $\times$10$^{13}$ & \citet{Rivilla2021a} \\
Vinyl amine  & C$_2$H$_3$NH$_2$ & 2021    & 8 & 4.5(0.6) $\times$10$^{13}$ & \citet{Zeng2021} \\
$n$-propanol  & {\it Ga-n}-C$_3$H$_7$OH   & 2022 & 12 &  5.5(0.4) $\times$10$^{13}$ & \citet{Jimenez-Serra2022} \\ 
$n$-propanol   & {\it Aa-n}-C$_3$H$_7$OH  & 2022 & 12 & 3.4(0.3) $\times$10$^{13}$ & \citet{Jimenez-Serra2022} \\
Phosphorus oxide cation    & PO$^+$ & 2022 & 2 &  6.0(0.7) $\times$10$^{11}$ & \citet{Rivilla2022_PO+} \\
1,2-ethenediol   & Z-(CHOH)$_2$   & 2022  & 8 &   1.8(0.1) $\times$10$^{13}$ & \citet{Rivilla2022a} \\
Isobutene  & CH$_3$CCH$_2$CH$_3$ & 2023 & 12 &    2.6(0.4) $\times$10$^{14}$ & \citet{Fatima2023} \\
Glycolamide    & $syn$-NH$_2$C(O)CH$_2$OH & 2023 & 9 &   7.4(0.7) $\times$10$^{12}$ & \citet{Rivilla2023} \\
Silacyclopropynylidene   & SiC$_2$  & 2023 & 3 &   1.02(0.04) $\times$10$^{13}$ & \citet{Massalkhi2023} \\
Carbonic acid  & cis-trans HOCOOH & 2023 & 6 &    6.4(0.4) $\times$10$^{12}$ & \citet{Sanz-Novo2023} \\
N-cyanomethanimine & CH$_2$NCN   & 2024 & 6 &   2.9(0.1) $\times$10$^{12}$ & \citet{SanAndres2024} \\
Protonated carbonyl sulphide    & HOCS$^+$  & 2024 & 4 &   9(2) $\times$10$^{12}$ & \citet{Sanz-Novo2024a} \\
Thionylimide    & $cis$-HNSO  & 2024 & 4 &   8(1) $\times$10$^{13}$ & \citet{Sanz-Novo2024b} \\
Magnesium sulfide & MgS  & 2024 & 2 &   2.0(0.5) $\times$10$^{10}$ & \citet{Rey-Montejo2024} \\
Sodium sulfide    &  NaS & 2024 & 2 &   5.0(1.1) $\times$10$^{10}$ & \citet{Rey-Montejo2024} \\
Dimethyl sulfide    &  CH$_3$SCH$_3$ & 2025 & 9 &   2.6(0.2) $\times$10$^{13}$ & \citet{sanz-novo2025} \\
Methyl formate    & trans-CH$_3$OCHO  & 2025   & 8 &  8.3(0.4) $\times$10$^{12}$ & \citet{Sanz-novo2025_mf} \\
N-methyl formamide    & cis N-CH$_3$NHCHO & 2025   & 9 &  1.5(0.1) $\times$10$^{13}$ & \citet{zeng2025} \\
2,5-cyclohexadien-1-thione    & c-C$_6$H$_6$S  &  2026 & 12  &  5.6(0.3) $\times$10$^{12}$ & \citet{Araki2026} \\
Lactaldehyde    & CH$_3$CHOHCOH & 2026   & 11 & 1.1(0.13) $\times$10$^{13}$ & \citet{sanz-novo2026} \\
Methoxyacetaldehyde    & CH$_3$OCH$_2$COH & 2026   & 11 & 0.33(0.03) $\times$10$^{13}$ & \citet{sanz-novo2026} \\    
    \midrule
    \end{tabular}}
\tabnote{\textit{Notes}: [a] Species reported before IAUS 405 (May 2026)
; [b] To compute the molecular abundances with respect to H$_{2}$, the value of the $N$ reported here needs to be divided by $N$(H$_2$)=1.35$\times$10$^{23}$ cm$^{-2}$, from \citet{martin2008};
[c] Column densities from \citet{SanAndres2024}.
}
\label{tab:detections}
\end{table}

\clearpage
\vspace{-2mm}

\bibliographystyle{aa}
\bibliography{biblio}

@ARTICLE{Bizzocchi2020,
       author = {{Bizzocchi}, L. and {Prudenzano}, D. and {Rivilla}, V.~M. and {Pietropolli-Charmet}, A. and {Giuliano}, B.~M. and {Caselli}, P. and {Mart{\'\i}n-Pintado}, J. and {Jim{\'e}nez-Serra}, I. and {Mart{\'\i}n}, S. and {Requena-Torres}, M.~A. and {Rico-Villas}, F. and {Zeng}, S. and {Guillemin}, J. -C.},
        title = "{Propargylimine in the laboratory and in space: millimetre-wave spectroscopy and its first detection in the ISM}",
      journal = {\aap},
         year = 2020,
        month = aug,
       volume = {640},
          eid = {A98},
        pages = {A98},
          doi = {10.1051/0004-6361/202038083},
archivePrefix = {arXiv},
       eprint = {2006.08401},
 primaryClass = {astro-ph.GA},
       adsurl = {https://ui.adsabs.harvard.edu/abs/2020A&A...640A..98B}
}

@ARTICLE{Colzi2024,
       author = {{Colzi}, L. and {Mart{\'\i}n-Pintado}, J. and {Zeng}, S. and {Jim{\'e}nez-Serra}, I. and {Rivilla}, V.~M. and {Sanz-Novo}, M. and {Mart{\'\i}n}, S. and {Zhang}, Q. and {Lu}, X.},
        title = "{Excitation and spatial study of a prestellar cluster towards G+0.693-0.027 in the Galactic centre}",
      journal = {\aap},
         year = 2024,
        month = oct,
       volume = {690},
          eid = {A121},
        pages = {A121},
          doi = {10.1051/0004-6361/202451382},
       adsurl = {https://ui.adsabs.harvard.edu/abs/2024A&A...690A.121C}
}

@ARTICLE{Fatima2023,
       author = {{Fatima}, Mariyam and {M{\"u}ller}, Holger S.~P. and {Zingsheim}, Oliver and {Lewen}, Frank and {Rivilla}, V{\'\i}ctor M. and {Jim{\'e}nez-Serra}, Izaskun and {Mart{\'\i}n-Pintado}, Jes{\'u}s and {Schlemmer}, Stephan},
        title = "{Millimetre and submillimetre spectroscopy of isobutene and its detection in the molecular cloud G+0.693}",
      journal = {\aap},
         year = 2023,
        month = dec,
       volume = {680},
          eid = {A25},
        pages = {A25},
          doi = {10.1051/0004-6361/202347112},
archivePrefix = {arXiv},
       eprint = {2309.17236},
 primaryClass = {astro-ph.GA},
       adsurl = {https://ui.adsabs.harvard.edu/abs/2023A&A...680A..25F}
}

@ARTICLE{Jimenez-Serra2022,
       author = {{Jim{\'e}nez-Serra}, Izaskun and {Rodr{\'i}guez-Almeida}, Lucas F. and {Mart{\'i}n-Pintado}, Jes{\'u}s and {Rivilla}, V{\'i}ctor M. and {Melosso}, Mattia and {Zeng}, Shaoshan and {Colzi}, Laura and {Kawashima}, Yoshiyuki and {Hirota}, Eizi and {Puzzarini}, Cristina and {Tercero}, Bel{\'e}n and {de Vicente}, Pablo and {Rico-Villas}, Fernando and {Requena-Torres}, Miguel A. and {Mart{\'i}n}, Sergio},
        title = "{Precursors of fatty alcohols in the ISM: Discovery of n-propanol}",
      journal = {\aap},
         year = 2022,
        month = jul,
       volume = {663},
          eid = {A181},
        pages = {A181},
          doi = {10.1051/0004-6361/202142699},
archivePrefix = {arXiv},
       eprint = {2204.08267},
 primaryClass = {astro-ph.GA},
       adsurl = {https://ui.adsabs.harvard.edu/abs/2022A&A...663A.181J}
}

@ARTICLE{Martin2008,
       author = {{Mart{\'i}n}, Sergio and {Requena-Torres}, M.~A. and {Mart{\'i}n-Pintado}, J. and {Mauersberger}, R.},
        title = "{Tracing Shocks and Photodissociation in the Galactic Center Region}",
      journal = {\apj},
         year = 2008,
        month = may,
       volume = {678},
       number = {1},
        pages = {245-254},
          doi = {10.1086/533409},
archivePrefix = {arXiv},
       eprint = {0801.3614},
 primaryClass = {astro-ph},
       adsurl = {https://ui.adsabs.harvard.edu/abs/2008ApJ...678..245M}
}

@ARTICLE{Massalkhi2023,
       author = {{Massalkhi}, S. and {Jim{\'e}nez-Serra}, I. and {Mart{\'\i}n-Pintado}, J. and {Rivilla}, V.~M. and {Colzi}, L. and {Zeng}, S. and {Mart{\'\i}n}, S. and {Tercero}, B. and {de Vicente}, P. and {Requena-Torres}, M.~A.},
        title = "{The first detection of SiC$_{2}$ in the interstellar medium}",
      journal = {\aap},
         year = 2023,
        month = oct,
       volume = {678},
          eid = {A45},
        pages = {A45},
          doi = {10.1051/0004-6361/202346822},
archivePrefix = {arXiv},
       eprint = {2308.09459},
 primaryClass = {astro-ph.GA},
       adsurl = {https://ui.adsabs.harvard.edu/abs/2023A&A...678A..45M}
}

@ARTICLE{McGuire2022,
       author = {{McGuire}, Brett A.},
        title = "{2021 Census of Interstellar, Circumstellar, Extragalactic, Protoplanetary Disk, and Exoplanetary Molecules}",
      journal = {\apjs},
         year = 2022,
        month = apr,
       volume = {259},
       number = {2},
          eid = {30},
        pages = {30},
          doi = {10.3847/1538-4365/ac2a48},
archivePrefix = {arXiv},
       eprint = {2109.13848},
 primaryClass = {astro-ph.GA},
       adsurl = {https://ui.adsabs.harvard.edu/abs/2022ApJS..259...30M}
}

@ARTICLE{Melosso2022,
       author = {{Melosso}, Mattia and {Bizzocchi}, Luca and {Gazzeh}, Houda and {Tonolo}, Francesca and {Guillemin}, Jean-Claude and {Alessandrini}, Silvia and {Rivilla}, V{\'\i}ctor M. and {Dore}, Luca and {Barone}, Vincenzo and {Puzzarin}, Cristina},
        title = "{Gas-phase identification of ( Z )-1,2-ethenediol, a key prebiotic intermediate in the formose reaction}",
      journal = {Chemical Communications},
         year = 2022,
        month = feb,
       volume = {58},
       number = {16},
        pages = {2750-2753},
       adsurl = {https://ui.adsabs.harvard.edu/abs/2022ChCom..58.2750M}
}

@ARTICLE{Requena-Torres2008,
       author = {{Requena-Torres}, M.~A. and {Mart{\'\i}n-Pintado}, J. and {Mart{\'\i}n}, S. and {Morris}, M.~R.},
        title = "{The Galactic Center: The Largest Oxygen-bearing Organic Molecule Repository}",
      journal = {\apj},
         year = 2008,
        month = jan,
       volume = {672},
       number = {1},
        pages = {352-360},
          doi = {10.1086/523627},
archivePrefix = {arXiv},
       eprint = {0709.0542},
 primaryClass = {astro-ph},
       adsurl = {https://ui.adsabs.harvard.edu/abs/2008ApJ...672..352R}
}

@ARTICLE{Rey-Montejo2024,
       author = {{Rey-Montejo}, M. and {Jimenez-Serra}, I. and {Martin-Pintado}, J. and {Rivilla}, V.~M. and {Megias}, A. and {San Andres}, D. and {Sanz-Novo}, M. and {Colzi}, L. and {Zeng}, S. and {Lopez-Gallifa}, A. and {Martinez-Henares}, A. and {Martin}, S. and {Tercero}, B. and {de Vicente}, P. and {Requena-Torres}, M.},
        title = "{Discovery of MgS and NaS in the Interstellar Medium and tentative detection of CaO}",
      journal = {arXiv e-prints},
         year = 2024,
        month = jul,
          eid = {arXiv:2407.07693},
        pages = {arXiv:2407.07693},
          doi = {10.48550/arXiv.2407.07693},
archivePrefix = {arXiv},
       eprint = {2407.07693},
 primaryClass = {astro-ph.GA},
       adsurl = {https://ui.adsabs.harvard.edu/abs/2024arXiv240707693R}
}

@ARTICLE{Rivilla2019,
       author = {{Rivilla}, V.~M. and {Mart{\'i}n-Pintado}, J. and {Jim{\'e}nez-Serra}, I. and {Zeng}, S. and {Mart{\'i}n}, S. and {Armijos-Abenda{\~n}o}, J. and {Requena-Torres}, M.~A. and {Aladro}, R. and {Riquelme}, D.},
        title = "{Abundant Z-cyanomethanimine in the interstellar medium: paving the way to the synthesis of adenine}",
      journal = {\mnras},
         year = 2019,
        month = feb,
       volume = {483},
       number = {1},
        pages = {L114-L119},
          doi = {10.1093/mnrasl/sly228},
archivePrefix = {arXiv},
       eprint = {1811.12862},
 primaryClass = {astro-ph.GA},
       adsurl = {https://ui.adsabs.harvard.edu/abs/2019MNRAS.483L.114R}
}

@ARTICLE{Rivilla2020b,
       author = {{Rivilla}, V{\'\i}ctor M. and {Mart{\'\i}n-Pintado}, Jes{\'u}s and {Jim{\'e}nez-Serra}, Izaskun and {Mart{\'i}n}, Sergio and {Rodr{\'i}guez-Almeida}, Lucas F. and {Requena-Torres}, Miguel A. and {Rico-Villas}, Fernando and {Zeng}, Shaoshan and {Briones}, Carlos},
        title = "{Prebiotic Precursors of the Primordial RNA World in Space: Detection of NH$_{2}$OH}",
      journal = {\apjl},
         year = 2020,
        month = aug,
       volume = {899},
       number = {2},
          eid = {L28},
        pages = {L28},
          doi = {10.3847/2041-8213/abac55},
archivePrefix = {arXiv},
       eprint = {2008.00228},
 primaryClass = {astro-ph.GA},
       adsurl = {https://ui.adsabs.harvard.edu/abs/2020ApJ...899L..28R}
}

@ARTICLE{Rivilla2021a,
       author = {{Rivilla}, V{\'i}ctor M. and {Jim{\'e}nez-Serra}, Izaskun and {Mart{\'i}n-Pintado}, Jes{\'u}s and {Briones}, Carlos and {Rodr{\'i}guez-Almeida}, Lucas F. and {Rico-Villas}, Fernando and {Tercero}, Bel{\'e}n and {Zeng}, Shaoshan and {Colzi}, Laura and {de Vicente}, Pablo and {Mart{\'i}n}, Sergio and {Requena-Torres}, Miguel A.},
        title = "{Discovery in space of ethanolamine, the simplest phospholipid head group}",
      journal = {Proceedings of the National Academy of Science},
         year = 2021,
        month = jun,
       volume = {118},
       number = {22},
          eid = {e2101314118},
        pages = {e2101314118},
          doi = {10.1073/pnas.2101314118},
archivePrefix = {arXiv},
       eprint = {2105.11141},
 primaryClass = {astro-ph.GA},
       adsurl = {https://ui.adsabs.harvard.edu/abs/2021PNAS..11801314R}
}

@ARTICLE{Rivilla2021b,
       author = {{Rivilla}, V.~M. and {Jim{\'e}nez-Serra}, I. and {Garc{\'i}a de la Concepci{\'o}n}, J. and {Mart{\'i}n-Pintado}, J. and {Colzi}, L. and {Rodr{\'i}guez-Almeida}, L.~F. and {Tercero}, B. and {Rico-Villas}, F. and {Zeng}, S. and {Mart{\'i}n}, S. and {Requena-Torres}, M.~A. and {de Vicente}, P.},
        title = "{Detection of the cyanomidyl radical (HNCN): a new interstellar species with the NCN backbone}",
      journal = {\mnras},
         year = 2021,
        month = sep,
       volume = {506},
       number = {1},
        pages = {L79-L84},
          doi = {10.1093/mnrasl/slab074},
archivePrefix = {arXiv},
       eprint = {2106.09652},
 primaryClass = {astro-ph.GA},
       adsurl = {https://ui.adsabs.harvard.edu/abs/2021MNRAS.506L..79R}
}

@ARTICLE{Rivilla2022a,
       author = {{Rivilla}, V{\'i}ctor M. and {Colzi}, Laura and {Jim{\'e}nez-Serra}, Izaskun and {Mart{\'i}n-Pintado}, Jes{\'u}s and {Meg{\'i}as}, Andr{\'e}s and {Melosso}, Mattia and {Bizzocchi}, Luca and {L{\'o}pez-Gallifa}, {\'A}lvaro and {Mart{\'i}nez-Henares}, Antonio and {Massalkhi}, Sarah and {Tercero}, Bel{\'e}n and {de Vicente}, Pablo and {Guillemin}, Jean-Claude and {Garc{\'i}a de la Concepci{\'o}n}, Juan and {Rico-Villas}, Fernando and {Zeng}, Shaoshan and {Mart{\'i}n}, Sergio and {Requena-Torres}, Miguel A. and {Tonolo}, Francesca and {Alessandrini}, Silvia and {Dore}, Luca and {Barone}, Vincenzo and {Puzzarini}, Cristina},
        title = "{Precursors of the RNA World in Space: Detection of (Z)-1,2-ethenediol in the Interstellar Medium, a Key Intermediate in Sugar Formation}",
      journal = {\apjl},
         year = 2022,
        month = apr,
       volume = {929},
       number = {1},
          eid = {L11},
        pages = {L11},
          doi = {10.3847/2041-8213/ac6186},
archivePrefix = {arXiv},
       eprint = {2203.14728},
 primaryClass = {astro-ph.GA},
       adsurl = {https://ui.adsabs.harvard.edu/abs/2022ApJ...929L..11R}
}

@ARTICLE{Rivilla2022_PO+,
       author = {{Rivilla}, V{\'i}ctor M. and {Garc{\'i}a De La Concepci{\'o}n}, Juan and {Jim{\'e}nez-Serra}, Izaskun and {Mart{\'i}n-Pintado}, Jes{\'u}s and {Colzi}, Laura and {Tercero}, Bel{\'e}n and {Meg{\'i}as}, Andr{\'e}s and {L{\'o}pez-Gallifa}, {\'A}lvaro and {Mart{\'i}nez-Henares}, Antonio and {Massalkhi}, Sara and {Mart{\'i}n}, Sergio and {Zeng}, Shaoshan and {De Vicente}, Pablo and {Rico-Villas}, Fernando and {Requena-Torres}, Miguel A. and {Cosentino}, Giuliana},
        title = "{Ionize Hard: Interstellar PO+ Detection}",
      journal = {Frontiers in Astronomy and Space Sciences},
         year = 2022,
        month = apr,
       volume = {9},
          eid = {829288},
        pages = {829288},
          doi = {10.3389/fspas.2022.829288},
archivePrefix = {arXiv},
       eprint = {2202.13928},
 primaryClass = {astro-ph.GA},
       adsurl = {https://ui.adsabs.harvard.edu/abs/2022FrASS...9.9288R}
}

@ARTICLE{Rivilla2022_nitriles,
       author = {{Rivilla}, V{\'\i}ctor M. and {Jim{\'e}nez-Serra}, Izaskun and {Mart{\'\i}n-Pintado}, Jes{\'u}s and {Colzi}, Laura and {Tercero}, Bel{\'e}n and {de Vicente}, Pablo and {Zeng}, Shaoshan and {Mart{\'\i}n}, Sergio and {Garc{\'\i}a de la Concepci{\'o}n}, Juan and {Bizzocchi}, Luca and {Melosso}, Mattia and {Rico-Villas}, Fernando and {Requena-Torres}, Miguel A.},
        title = "{Molecular Precursors of the RNA-World in Space: New Nitriles in the G+0.693{\ensuremath{-}}0.027 Molecular Cloud}",
      journal = {Frontiers in Astronomy and Space Sciences},
         year = 2022,
        month = jul,
       volume = {9},
          eid = {876870},
        pages = {876870},
          doi = {10.3389/fspas.2022.876870},
archivePrefix = {arXiv},
       eprint = {2206.01053},
 primaryClass = {astro-ph.GA},
       adsurl = {https://ui.adsabs.harvard.edu/abs/2022FrASS...9.6870R}
}

@ARTICLE{Rivilla2023,
       author = {{Rivilla}, V{\'\i}ctor M. and {Sanz-Novo}, Miguel and {Jim{\'e}nez-Serra}, Izaskun and {Mart{\'\i}n-Pintado}, Jes{\'u}s and {Colzi}, Laura and {Zeng}, Shaoshan and {Meg{\'\i}as}, Andr{\'e}s and {L{\'o}pez-Gallifa}, {\'A}lvaro and {Mart{\'\i}nez-Henares}, Antonio and {Massalkhi}, Sarah and {Tercero}, Bel{\'e}n and {de Vicente}, Pablo and {Mart{\'\i}n}, Sergio and {San Andr{\'e}s}, David and {Requena-Torres}, Miguel A. and {Alonso}, Jos{\'e} Luis},
        title = "{First Glycine Isomer Detected in the Interstellar Medium: Glycolamide (NH$_{2}$C(O)CH$_{2}$OH)}",
      journal = {\apjl},
         year = 2023,
        month = aug,
       volume = {953},
       number = {2},
          eid = {L20},
        pages = {L20},
          doi = {10.3847/2041-8213/ace977},
archivePrefix = {arXiv},
       eprint = {2307.11507},
 primaryClass = {astro-ph.GA},
       adsurl = {https://ui.adsabs.harvard.edu/abs/2023ApJ...953L..20R}
}

\end{document}